\documentclass[conference]{IEEEtran}
\IEEEoverridecommandlockouts
\usepackage{cite}
\usepackage{amsmath}
\usepackage{graphicx}
\usepackage{booktabs}
\usepackage{xcolor}
\usepackage{url}
\usepackage[T1]{fontenc}

\newcommand{\sass}[1]{\texttt{#1}}
\newcommand{\reliable}{\sass{.RELIABLE}}
\newcommand{\reconv}{\sass{.RECONVERGENT}}

\begin{document}

\title{Characterizing Warp Divergence from Pascal to Blackwell}

\author{\IEEEauthorblockN{Alpin Dale}
\IEEEauthorblockA{alpin@dphn.ai}}

\maketitle

\begin{abstract}
Since the Volta architecture introduced Independent Thread Scheduling (ITS) in
2017, giving every thread its own program counter, NVIDIA GPUs have been widely
assumed to handle control-flow divergence in a fixed manner. We test that
assumption with a cross-generational study of warp divergence
spanning Ampere, Hopper, and both the datacenter and consumer variants of
Blackwell, anchored against a pre-ITS Pascal baseline. Combining cycle-accurate
microbenchmarks, hardware performance
counters, and static analysis of compiler-generated SASS, we separate what has
remained invariant from what has evolved. We find that the dynamic cost of
divergence varies little across the tested generations. Divergent paths serialize
strictly linearly in the number of paths $k$ ($T(k)\approx s\!\cdot\!k$, with no
super-linear reconvergence penalty), warp execution efficiency falls as exactly
$32/k$, and this behavior is independent of occupancy and identical across the
four post-ITS generations; the same linear cost, predication remedy, and
occupancy-invariance already hold on the pre-ITS Pascal baseline, so this
behavior predates ITS. The static reconvergence machinery the compiler emits has
changed substantially. Pascal reconverges through a per-warp instruction stack
(\sass{SSY}/\sass{SYNC}) that the ITS generations replace with barrier-register
instructions. The fraction of divergent branches that reconverge later
than the immediate post-dominator, the deferred reconvergence that classical
SIMT-stack hardware relied on, collapses from $29$ cases on Ampere to $2$ on
Blackwell. Blackwell additionally introduces a two-tier convergence
barrier (a previously undocumented \reliable{} class distinct from \reconv{},
encoded in a 2-bit field), uniform-branch instructions, and explicit partial-mask
warp synchronization, none of which exist on Ampere or Hopper. Controlled bit-flip
experiments show this barrier class is a static compiler classification with no
observable runtime effect in our tests. ITS has changed
beneath a stable programmer-visible cost model. Divergence still serializes
predictably, while the control-flow ISA and reconvergence mechanisms continue to
evolve.
\end{abstract}


\section{Introduction}
The single-instruction, multiple-thread (SIMT) execution model groups 32 threads
into a warp that shares one instruction-fetch stream~\cite{lindholm2008tesla}.
When threads in a warp take different sides of a data-dependent branch, the warp
diverges. The hardware must execute the taken paths separately, leaving lanes
idle on each path and eroding the SIMD efficiency that makes GPUs fast. Control-flow
divergence has accordingly been one of the most studied GPU performance
pathologies, motivating a long line of hardware reconvergence
mechanisms~\cite{fung2007dwf,fung2011tbc,meng2010dws,diamos2011frontiers} and
compiler techniques~\cite{coutinho2011divergence,sampaio2013divergence,lee2013scalarization,saumya2022darm}.

Pre-Volta GPUs managed divergence with a per-warp reconvergence stack that forced
threads back together at the immediate post-dominator (IPDom) of each
branch~\cite{fung2007dwf}. The model gives a correct reconvergence rule, but it
can block a thread holding a lock while its peers spin, producing
deadlocks~\cite{habermaier2012correctness}. Volta's Independent Thread
Scheduling (ITS) gave each thread an independent program counter and call stack to
break this limitation~\cite{nvidia2017volta}. Because ITS is described only
informally in vendor whitepapers and its hardware is closed, a widespread
assumption has taken hold. Post-Volta divergence handling is often treated as
essentially settled, with Ampere results presumed to transfer unchanged to Hopper
or Blackwell.

Recent microarchitecture-dissection papers reinforce this blind spot. Thorough
microbenchmark studies of Volta~\cite{jia2018volta}, Turing~\cite{jia2019turing},
Ampere~\cite{abdelkhalik2022ampere}, Hopper~\cite{luo2024hopper}, and
Blackwell~\cite{jarmusch2025blackwell} characterize the memory hierarchy, tensor
cores, and instruction throughput in depth, while saying almost nothing about
control-flow divergence. The most detailed SASS-level reverse engineering of GPU
control flow to date~\cite{shoushtary2024controlflow} covers a single
architecture (Turing). No prior work asks whether ITS divergence behavior has
changed across generations, nor measures it on Blackwell.

We close that gap. Using microbenchmarks, GPU hardware performance counters, and
static analysis of compiler-emitted SASS, we characterize warp divergence on four
post-ITS GPUs spanning three generations and two Blackwell variants, together with
a pre-ITS Pascal baseline (Table~\ref{tab:testbed}). Our guiding question is whether ITS has stood still.
The answer has two parts. The dynamic behavior a programmer observes is
invariant, while the static control-flow machinery the compiler and ISA expose
has changed.

\noindent\textbf{Contributions.}
\begin{itemize}
\item A cross-generational cost model for warp divergence
(Ampere, Hopper, Blackwell $\times 2$), showing that divergence serializes
linearly ($T(k)\approx s\!\cdot\!k$) with a small constant per path and no
super-linear reconvergence penalty up to a full 32-way split; the same law already
governs a pre-ITS Pascal part, so linear serialization predates ITS
(Section~\ref{sec:cost}).
\item An independent hardware-counter confirmation that warp execution efficiency
falls as exactly $32/k$, generation-invariant, together with the finding that the
divergence penalty is occupancy-invariant because it is an instruction-issue cost
that additional resident warps cannot hide (Section~\ref{sec:cost}).
\item A cross-generational static analysis of reconvergence placement relative to
the IPDom, showing that deferred (later-than-IPDom) reconvergence, the regime
classical stack hardware depends on, all but disappears on Blackwell, and that the
pre-ITS part still uses the literal \sass{SSY}/\sass{SYNC} instruction stack that
the barrier-register scheme replaced (Section~\ref{sec:reconv}).
\item The first description of Blackwell's two-tier convergence-barrier scheme
(\reconv{} vs.\ \reliable{}, a 2-bit barrier class), its uniform-branch
instruction \sass{BRA.U}, and its explicit partial-mask \sass{WARPSYNC}, none of
which appear on Ampere or Hopper; controlled bit-flip experiments further show
the barrier class is a static classification with no observable runtime effect in
our tests (Section~\ref{sec:reconv}).
\end{itemize}

\section{Background}
\label{sec:bg}

\subsection{SIMT, reconvergence, and the IPDom stack}
A warp executes one instruction at a time across an active mask of up to 32
lanes. A divergent branch partitions the active lanes; the hardware executes each
resulting path with a reduced active mask and must eventually reconverge
them. The classical mechanism is a per-warp stack that reconverges threads at the
immediate post-dominator of the branch, the earliest instruction that all paths
must reach~\cite{fung2007dwf}. A large body of work improves on IPDom-stack
reconvergence through dynamic warp formation~\cite{fung2007dwf}, thread-block
compaction~\cite{fung2011tbc}, warp subdivision~\cite{meng2010dws}, thread
frontiers~\cite{diamos2011frontiers}, stack-less designs~\cite{collange2011stackless},
and divergence-aware scheduling~\cite{rogers2013daws,narasiman2011largewarp,vaidya2013compaction}.

\subsection{Independent Thread Scheduling}
Volta replaced the single per-warp PC with per-thread state, enabling threads on
different paths to be scheduled independently and guaranteeing forward
progress~\cite{nvidia2017volta}. This removed the classic spin-lock deadlock that
the stack model permits~\cite{habermaier2012correctness,eltantawy2016mimd}, at the
cost of no longer guaranteeing implicit reconvergence, hence the introduction of
explicit \sass{\_\_syncwarp()} convergence barriers. How ITS schedules divergent
paths in practice, and whether that scheduling has changed since Volta, is not
publicly documented.

\subsection{SASS control flow and the scheduling control word}
NVIDIA GPUs execute a proprietary binary ISA (SASS). Since Volta, each instruction
is 128 bits and carries a compiler-assigned scheduling control word
encoding a stall count, a yield hint, scoreboard dependencies, and a reuse
cache~\cite{jia2018volta}. Post-Volta control flow is managed by explicit
instructions. \sass{BSSY} establishes a reconvergence barrier at a target PC, and
\sass{BSYNC} waits on it rather than relying on an implicit hardware
stack~\cite{shoushtary2024controlflow}. We analyze how these instructions are
emitted across generations in Section~\ref{sec:reconv}.

\section{Experimental Setup}
\label{sec:setup}
\begin{table}[t]
\centering
\caption{Testbed. The post-ITS GPUs run CUDA~13; the pre-ITS Pascal baseline runs CUDA~12.4.}
\label{tab:testbed}
\resizebox{\columnwidth}{!}{%
\begin{tabular}{@{}llrl@{}}
\toprule
Architecture & GPU & SMs & Compute cap. \\
\midrule
Pascal (pre-ITS)     & GTX 1080 & 20 & sm\_61 \\
Ampere    & RTX A6000 / RTX 3090 & 84/82 & sm\_86 \\
Hopper    & H100                 & 132   & sm\_90 \\
Blackwell (server)   & Jetson AGX Thor & 20 & sm\_110 \\
Blackwell (consumer) & RTX 5080        & 84 & sm\_120 \\
\bottomrule
\end{tabular}}
\end{table}

We characterize divergence along two axes. Dynamic experiments in
Section~\ref{sec:cost} run microbenchmarks on the tested GPUs. Static
experiments in Section~\ref{sec:reconv} compile a kernel corpus for every
architecture and analyze the resulting SASS. Because compilation targets any
architecture from one host, the static study additionally covers the full sm\_80
through sm\_120 range; the pre-ITS Pascal (sm\_61) SASS was produced on the Pascal
host itself, as current toolchains no longer target it.

\noindent\textbf{Cycle measurements.} The dynamic microbenchmarks bracket the
region of interest with the \sass{clock64()} SM cycle counter. Cycle counts are
frequency-independent, so per-path slopes are directly comparable across
GPUs despite different clocks. Each datum is the median of 201 launches. The
coefficient of variation is below $10^{-4}$ on every GPU, so the measurement is
effectively deterministic and we omit error bars.

\noindent\textbf{Forcing genuine divergence.} A subtle methodological hazard is
that the compiler may predicate a short or simple branch, executing both
sides under a per-lane mask with the full warp active, rather than emit a true
divergent branch. Predicated code has the same instruction count but does not
reduce SIMD utilization, so it could be mistaken for divergence in a timing
experiment while a counter would show full efficiency. We defend against this in
two ways. Every divergent kernel forces a real branch, using a long,
non-mergeable loop body per path bracketed by per-class \sass{\_\_syncwarp()},
and the disassembly confirms that each kernel emits \sass{BSSY}/\sass{BSYNC} reconvergence
barriers (\sass{SSY}/\sass{SYNC} on the pre-ITS part). We additionally validate our efficiency counter against kernels of known
divergence (a convergent kernel reads $32.0$ active threads per instruction, a
2-way split $16.3$, a 32-way split $1.6$).

\noindent\textbf{Efficiency counter.} We measure warp execution efficiency with
the Nsight Compute metric \texttt{smsp\_\_thread\_inst\_executed\_per\_}\allowbreak\texttt{inst\_executed.ratio},
the average number of active threads per issued
warp-instruction (32 when fully converged).

\section{The Cost of Divergence}
\label{sec:cost}

\subsection{Divergence is linear path serialization}
\begin{figure}[t]
\centering
\includegraphics[width=\columnwidth]{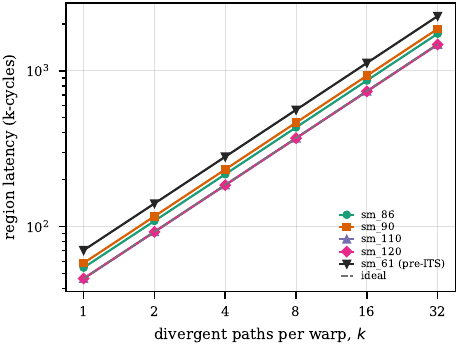}
\caption{Divergence cost is linear in the number of paths $k$ on every
generation, including the pre-ITS Pascal part. All curves are parallel to the
ideal $k{\times}$ serialization line; the per-path slope is the only difference,
and the two Blackwell parts are indistinguishable. Log--log axes.}
\label{fig:cost}
\end{figure}

Our central dynamic experiment varies the number of distinct paths $k$ a warp
takes through a balanced divergent region. All $k$ paths perform identical work, so
an ideal SIMT machine that serializes paths should take time proportional to $k$.
Figure~\ref{fig:cost} plots the measured region latency against $k$ for all four
post-ITS GPUs and the pre-ITS Pascal part on log--log axes.

The data fit $T(k) \approx s\!\cdot\!k + c$ with high precision. The per-path slope
$s$ is $54.1$\,k cycles on Ampere, $58.1$\,k on Hopper, $46.1$\,k on
both Blackwell parts, and $70.1$\,k on the pre-ITS Pascal part. The constant
overhead $c$ is roughly $2$\,k cycles on the post-ITS parts and negligible on
Pascal. A full 32-way split therefore costs $31.7$--$31.9\times$ a single path on
every part, i.e.,
essentially perfect linear serialization with no measurable super-linear
reconvergence penalty. The cost model is generation-invariant, with divergence
serializing the same way on Ampere as on Blackwell. Crucially, the pre-ITS Pascal
part obeys the identical linear law (per-path cost flat to within $0.6\%$ across
$k$), so linear path serialization is a property of the SIMT execution model
itself rather than of ITS; Pascal's larger per-path constant merely reflects its
lower single-thread throughput. The per-path constant tracks
general single-thread throughput. Blackwell executes a path about 15--20\%
faster than Hopper, rather than changing how divergence is handled. The
numerically identical slopes of
the server (sm\_110) and consumer (sm\_120) Blackwell parts indicate that the
per-path cost is an architecture-family property, not a chip-specific one.

\subsection{Efficiency falls as $32/k$}
\begin{figure}[t]
\centering
\includegraphics[width=\columnwidth]{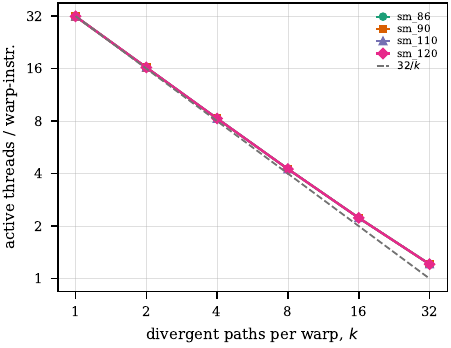}
\caption{Warp execution efficiency (active threads per issued instruction) falls
as $32/k$ on every generation. The four curves overlap. Measured with Nsight
Compute. Log--log axes.}
\label{fig:eff}
\end{figure}

The timing result implies that SIMD efficiency should fall as $32/k$. A $k$-way
divergent region issues the same instructions $k$ times, each with $32/k$ active
lanes. Figure~\ref{fig:eff} confirms this directly with hardware counters. On all
four post-ITS GPUs the average number of active threads per warp-instruction tracks the
$32/k$ ideal almost exactly ($31.9, 16.2, 8.3, 4.25, 2.23, 1.21$ for
$k=1,2,4,8,16,32$), and the four curves are indistinguishable. This
independent instrument confirms that the linear cost of Figure~\ref{fig:cost} is
genuine divergence (lost SIMD utilization) rather than an artifact of predicated
or redundant execution. The pre-ITS Pascal part is omitted from this counter
measurement, as its performance counters are gated to administrators on the
available driver; its linear cost (Fig.~\ref{fig:cost}) and occupancy-invariant
penalty (Fig.~\ref{fig:occ}) establish the same lost-utilization effect by timing.

\subsection{Predication removes the serialization}
Replacing a two-way divergent branch with a predicated, branch-free formulation
that computes both arms collapses the cost from $2.00\times$ to $1\times$ on all
four post-ITS GPUs (measured ratios within $0.5\%$ of $2.00$), and from
$1.94\times$ on the pre-ITS Pascal part. This quantifies the
well-known guidance to predicate small branches, and confirms it is equally valid
from Pascal through Blackwell.

\subsection{The penalty is occupancy-invariant}
\begin{figure}[t]
\centering
\includegraphics[width=\columnwidth]{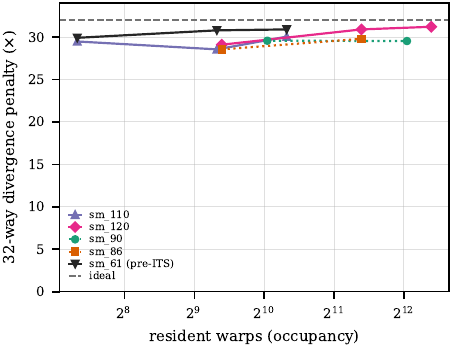}
\caption{The 32-way divergence penalty is occupancy-invariant: it stays near
$30\times$ from light to oversubscribed occupancy, on the pre-ITS Pascal part as
well as the post-ITS generations. Dotted series mark the two generations measured
at fewer occupancy points. Divergence is an issue-rate cost that resident warps
cannot hide.}
\label{fig:occ}
\end{figure}

One might expect a heavily occupied GPU to hide divergence by interleaving
other warps. Figure~\ref{fig:occ} shows otherwise. It sweeps the number of resident warps
from light to oversubscribed and measures the 32-way divergence penalty. The
penalty stays at $28$--$31\times$ across the entire range, including the pre-ITS
Pascal part ($29.9$--$30.9\times$), because divergence
raises the number of issued instructions. A divergent warp issues $k\times$ more
instructions, while occupancy hides latency rather than issue count. Divergence
therefore cannot be amortized by adding warps, which makes its
elimination (e.g., by predication or data reorganization) the only effective
remedy regardless of how busy the GPU is.

\subsection{Forward progress holds, but paths run in order}
ITS guarantees forward progress, and our experiments confirm it on every post-ITS
generation. A warp in which all 32 lanes contend for a single lock (each acquiring,
entering a critical section, and releasing) terminates correctly with all lanes
served on every generation we tested, the pre-ITS Pascal part included; this
particular lock releases within the acquiring branch, so it does not exercise the
classical stack deadlock, which requires an unbounded intra-warp peer-wait we do
not run on shared hardware. However, a
producer/consumer probe in which one lane must run before the others can proceed
reveals that divergent paths are scheduled in structural order and run to
completion. The consumer lanes observe the producer's result after at most two
spin iterations even when the producer performs $10^{7}$ iterations of work. ITS
thus provides the forward-progress guarantee without fine-grained interleaving of
divergent paths in the absence of contended synchronization, giving the model's
promise a practical scope.

\section{Reconvergence Across Generations}
\label{sec:reconv}
If the dynamic cost of divergence is stable, the static machinery the compiler
emits to manage it still changes. We compile a 38-kernel corpus spanning simple and
nested branches, switch statements, loops with early exits, short-circuit
evaluation, and deliberately irreducible control flow for every architecture from
sm\_80 to sm\_120, and analyze the resulting SASS. For each divergent branch on
these barrier-register generations we
reconstruct the control-flow graph, compute the immediate post-dominator, and
classify where the compiler places the matching \sass{BSYNC} reconvergence point;
the pre-ITS Pascal part, which has no such barriers, we treat separately first.

\subsection{From an instruction stack to barrier registers}
The pre-ITS Pascal part exposes the mechanism the later generations replaced. Its
divergent regions are bracketed by the classical reconvergence-stack instructions
\sass{SSY}, which pushes a reconvergence PC, and \sass{SYNC}, which pops it; nested
divergence produces nested \sass{SSY}/\sass{SYNC} pairs. Not a single \sass{BSSY},
\sass{BSYNC}, or \sass{BREAK} appears anywhere in the Pascal corpus. Every
post-Volta generation instead manages reconvergence with the barrier-register
instructions \sass{BSSY}/\sass{BSYNC} analyzed below, and Blackwell layers its
reliability-tagged two-tier scheme on top. The progression is thus a per-warp
instruction stack (Pascal) $\rightarrow$ bare barrier registers (Ampere, Hopper)
$\rightarrow$ reliability-tagged barrier registers (Blackwell), even as the
measured cost of divergence (Section~\ref{sec:cost}) stays fixed across all of
them.

\subsection{Deferred reconvergence disappears}
\begin{figure}[t]
\centering
\includegraphics[width=\columnwidth]{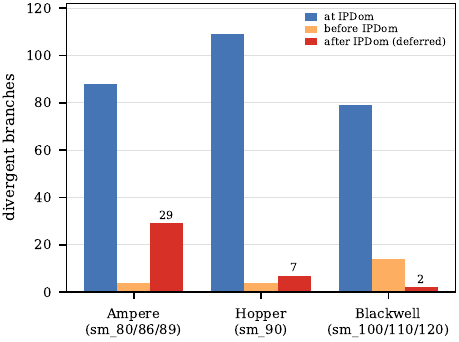}
\caption{Reconvergence placement relative to the immediate post-dominator across
generations. Deferred (later-than-IPDom) reconvergence, annotated, collapses from
$29$ on Ampere to $2$ on Blackwell.}
\label{fig:reconv}
\end{figure}

Figure~\ref{fig:reconv} shows the result. On Ampere, $29$ divergent branches
reconverge later than the IPDom. Deferred reconvergence is the regime in which
the classical stack postpones the merge through unrolled or nested loops.
This deferred tail shrinks to $7$ on Hopper and to just $2$ on Blackwell. The
fraction of branches reconverging exactly at the IPDom rises correspondingly from
$72.7\%$ (Ampere) to $90.8\%$ (Hopper), with Blackwell at $83.2\%$ trading the
vanished deferred tail for a new class of earlier-than-IPDom reconvergence
discussed below. Successive generations tighten reconvergence toward the
post-dominator and, in a controlled way, increasingly past it.

\subsection{A two-tier convergence barrier on Blackwell}
Blackwell's earlier-than-IPDom reconvergences are not ordinary barriers. On
sm\_100 and later, the \sass{BSSY}/\sass{BSYNC} instructions carry one of two
modifiers that do not exist on Hopper or earlier, where the barriers are bare. We
find these modifiers occupy a 2-bit reliability field in the instruction
encoding with two non-zero values:
\begin{itemize}
\item \reconv{}: the true post-dominator merge, the genuine reconvergence point,
which we observe is never broken out of; and
\item \reliable{}: an early, partial reconvergence placed at
unrolled-iteration and short-circuit-clause boundaries, nested inside the
enclosing \reconv{} region.
\end{itemize}
The distinction is visible in the encoding as well as in placement. The
\sass{BREAK} instruction is encoding-restricted to the \reliable{} class, so
\reliable{} is the
barrier class that a subset of lanes may exit early via \sass{BREAK}, while
\reconv{} is the merge that all lanes ultimately reach. This two-tier scheme, an
early partial-reconvergence barrier layered inside an IPDom-exact one, cannot be
expressed by the single bare barrier of earlier generations, and accounts for
Blackwell's controlled population of before-IPDom reconvergences.

We further test whether the field carries any runtime meaning beyond this encoding
role. We rewrite the 2-bit field directly in compiled cubins, verify the single
edit in the disassembly, and run the mutated kernels through the driver API. A
positive control, in which disabling a store changes the output, confirms that
the driver executes the patched bytes rather than re-deriving them.

With that control in place, flipping the field across all of its values leaves
output, completion, warp-level forward progress, a convergence-sensitive partner
exchange, and cycle counts unchanged on every kernel we tried. This includes
mismatched \sass{BSSY}/\sass{BSYNC} pairs and a \sass{BREAK.RECONVERGENT}
encoding that the disassembler can print after mutation but the assembler does
not emit for this control-flow pattern. We therefore read the field as a static
compiler and assembler classification rather than a runtime control, within the
tested executions. Reconvergence and forward-progress behavior is determined by
the barrier-register machinery and the \sass{BSSY}/\sass{BSYNC}/\sass{BREAK}
opcodes, while the reliability bit records which barrier is the true
post-dominator merge versus a breakable partial one. Blackwell also executes the
pre-Blackwell bare encoding identically in these tests.

\subsection{Uniform branches and partial-mask synchronization}
Two further control-flow features are Blackwell-only in our corpus
(Table~\ref{tab:isa}). First, Blackwell emits a uniform-branch instruction
\sass{BRA.U}, predicated on a uniform datapath register and preceded by a uniform
\sass{ISETP}, that the compiler uses to mark provably non-divergent branches
($23$ occurrences on Blackwell, $0$ on Ampere/Hopper); it replaces divergent
\sass{BRA}s where the compiler can prove uniformity, making the static
distinction between uniform and divergent control flow. Second, partial-mask warp
intrinsics (a \sass{\_\_shfl}/\sass{\_\_syncwarp} under a data-dependent mask) emit
an explicit \sass{WARPSYNC} on Blackwell but compile to bare shuffles with no
barrier on Ampere and Hopper. Blackwell makes partial-mask reconvergence explicit
in the ISA. Across all generations we never observe the \sass{BMOV} barrier-spill
instruction, and Blackwell caps the live reconvergence-barrier nesting depth at $3$
versus $4$ on older parts, predicating one additional level instead of spilling.

\begin{table}[t]
\centering
\caption{Control-flow ISA features by generation (static corpus). Blackwell adds
several mechanisms absent on Ampere/Hopper.}
\label{tab:isa}
\resizebox{\columnwidth}{!}{%
\begin{tabular}{@{}lccc@{}}
\toprule
Feature & Ampere & Hopper & Blackwell \\
\midrule
Reconverge at IPDom (rate)        & 72.7\% & 90.8\% & 83.2\% \\
Deferred (after-IPDom) branches   & 29     & 7      & 2 \\
Barrier modifiers (\reconv/\reliable) & none & none & \textbf{yes} \\
Uniform branch \sass{BRA.U}        & 0      & 0      & \textbf{23} \\
Partial-mask \sass{WARPSYNC}       & no     & no     & \textbf{yes} \\
Max barrier nesting depth          & 4      & 4      & 3 \\
\bottomrule
\end{tabular}}
\end{table}

\section{Discussion}
\label{sec:disc}
Our results draw a clean line between the parts of ITS that are stable and the
parts that are moving.

\noindent\textbf{For programmers and performance models}, the stability is good
news. A divergence cost model calibrated on one post-Volta generation transfers to
the next: $k$-way divergence costs $k\times$, efficiency is $32/k$, predication
recovers it, and no amount of occupancy hides it. Analytical
models~\cite{hong2009model} and simulators~\cite{bakhoda2009gpgpusim,khairy2020accelsim}
can treat the divergence-cost kernel as fixed across Ampere--Blackwell. The
occupancy-invariance result is a particularly actionable correction to the common
intuition that a busy GPU ``absorbs'' divergence; it does not, because divergence
inflates issue count rather than exposing latency.

\noindent\textbf{For architects and tool builders}, the evolving static machinery
is where the action is. The disappearance of deferred reconvergence and the
appearance of a two-tier barrier show that Blackwell's compiler and ISA represent
reconvergence distinctions more explicitly than earlier generations. Our
bit-flip experiments indicate that the \reliable{} field itself is not a runtime
control in the tested executions, but it remains concrete ISA-level evidence that
NVIDIA continues to expose new control-flow structure well after ITS. This
contradicts the assumption that the model froze with Volta. Binary-analysis tools~\cite{villa2019nvbit,stephenson2015sassi}
and SASS-level models~\cite{shoushtary2024controlflow,huerta2025gpucores} that
target a single generation will mis-model Blackwell control flow unless they
incorporate these instructions.

\noindent\textbf{Limitations.} Our dynamic study now includes a pre-ITS Pascal
part, but our bounded forward-progress probes complete on it as well; forcing the
classical stack deadlock needs an unbounded intra-warp peer-wait that we do not
run on shared hardware, so we still rely on prior
analysis~\cite{habermaier2012correctness} for that contrast. Our \reliable{}
result rests on single-warp bit-flip experiments on sm\_110, and does not exclude
effects in untested regimes such as specific multi-warp scheduling corners or
consumption of the field by profilers and debuggers. Finally, our divergence kernels are designed to
isolate path serialization. Real applications interleave divergence with memory
divergence, which we do not separate here.

\section{Related Work}
\label{sec:rel}
\textbf{Microarchitecture dissection.} A rich line of microbenchmark studies has
characterized successive NVIDIA generations, including Volta and
Turing~\cite{jia2018volta,jia2019turing},
Ampere~\cite{abdelkhalik2022ampere}, Hopper~\cite{luo2024hopper}, and
Blackwell~\cite{jarmusch2025blackwell}, as well as the core
microarchitecture and issue logic~\cite{huerta2025gpucores}. These works focus on
memory, tensor cores, and instruction throughput. Control-flow divergence is
essentially absent from them. We add the control-flow dimension and, uniquely,
study how it changes across generations rather than within one.

\noindent\textbf{Divergence mechanisms and mitigation.} The reconvergence problem
has a long history, including dynamic warp formation~\cite{fung2007dwf}, thread-block
compaction~\cite{fung2011tbc}, warp subdivision~\cite{meng2010dws}, large
warps~\cite{narasiman2011largewarp}, intra-warp compaction~\cite{vaidya2013compaction},
thread frontiers~\cite{diamos2011frontiers}, stack-less
reconvergence~\cite{collange2011stackless}, and divergence-aware
scheduling~\cite{rogers2013daws}, complemented by compiler divergence
analysis~\cite{coutinho2011divergence,sampaio2013divergence} and
scalarization/melding~\cite{lee2013scalarization,saumya2022darm}. This body of work
largely predates or assumes the IPDom-stack model. Our measurements show how far
production hardware has since moved its reconvergence points, and characterize the
forward-progress behavior that ITS~\cite{nvidia2017volta,eltantawy2016mimd}
introduced.

\noindent\textbf{SASS control flow.} The closest prior work reverse-engineers
Turing's control-flow instructions and builds a model of its reconvergence
mechanism~\cite{shoushtary2024controlflow}. We extend that style of analysis across
four architecture families (eight compute capabilities, sm\_61 through sm\_120),
contrasting the pre-ITS \sass{SSY}/\sass{SYNC} stack with the barrier-register
scheme, and surface the Blackwell-specific instructions and
barrier classes it could not have observed.

\noindent\textbf{Divergence cost measurement.} An early study benchmarked the cost
of thread divergence on pre-ITS hardware~\cite{bialas2015divergence}. We revisit
that question across four post-ITS generations and, closing the loop with that
pre-ITS setting, a Pascal part, with a hardware-counter cross-check and a control
for compiler predication, and contribute the occupancy-invariance and
cross-generational-invariance results.

\section{Conclusion}
\label{sec:conc}
We asked whether Independent Thread Scheduling has stood still since Volta. The
answer is a qualified no. The measured cost of divergence is stable across
Ampere, Hopper, and both Blackwell variants, and its linear form already appears
on pre-ITS Pascal. Divergent regions serialize
linearly, efficiency falls as $32/k$, predication recovers the lost utilization,
and additional occupancy does not hide the issue cost. At the same time, the
reconvergence machinery exposed by the compiler and ISA has changed markedly. The
pre-ITS part reconverges through a classical \sass{SSY}/\sass{SYNC} instruction
stack; every ITS generation replaces it with barrier registers,
deferred reconvergence has all but vanished, and Blackwell introduces a two-tier
\reconv{}/\reliable{} barrier scheme, uniform branches, and explicit partial-mask
synchronization that no earlier generation emits. Performance models can rely on
the measured cost behavior across these generations, while architecture and
tooling work must account for the changing control-flow ISA.

\bibliographystyle{IEEEtran}
\bibliography{references}

\end{document}